\documentclass[iop]{emulateapj}
\usepackage{graphicx}
\usepackage{epsfig}
\usepackage{epstopdf}
\usepackage{natbib}
\newcommand{\sdo}{{\em SDO{}}}

\newcommand{\iris}{{\em IRIS{}}}

\newcommand{\longacknowledgment}{\iris{} is a NASA small explorer developed and operated by LMSAL
with major contributions to downlink communications by ESA and 
Norwegian Space Centre. This work is supported by NASA (NNG09FA40C, \iris) and the UK Science and Technology Facilities Council and EU Horizon 2020 research programme (grant No. 647214).
The simulations were run on Pleiades (project s1061).}
\begin{document}

\shorttitle{Solar coronal heating associated with spicules}
\shortauthors{De Pontieu et al.}
\title{Observations and numerical models of solar coronal heating
  associated with spicules}
\author{B. De Pontieu\altaffilmark{1,2} \& I. De
  Moortel\altaffilmark{3} \& J. Martinez-Sykora\altaffilmark{1,4} \& S. W. McIntosh\altaffilmark{5}}
\affil{\altaffilmark{1}Lockheed Martin Solar \& Astrophysics Lab, Org.\ A021S, Bldg.\ 252, 3251 Hanover Street Palo Alto, CA~94304, USA}
\affil{\altaffilmark{2}Institute of Theoretical Astrophysics, University of Oslo, Blindern, P.O. Box 1029 Blindern, N-0315 Oslo, Norway}
\affil{\altaffilmark{3}School of Mathematics and Statistics, University of St Andrews, St Andrews, Fife, KY16 9SS, UK}
\affil{\altaffilmark{4}Bay Area Environmental Research Institute, Sonoma, CA~94952, USA}
\affil{\altaffilmark{5}High Altitude Observatory, National Center for Atmospheric Research, P.O. Box 3000, Boulder, CO 80307}

\begin{abstract}
Spicules have been proposed as significant contributors
to the mass and energy balance of the corona. While previous observations
have provided a glimpse of short-lived transient brightenings in the
corona that are associated with spicules, these observations have been
contested and are the subject of a vigorous debate both on the modeling
and the observational side. Therefore, it remains unclear whether plasma
is heated to coronal temperatures in association with spicules. 
We use high-resolution observations of the chromosphere and transition 
region with the Interface Region Imaging Spectrograph (\iris) and of
the corona with the Atmospheric Imaging Assembly (AIA) onboard the
Solar Dynamics Observatory (SDO) to show evidence of the formation of 
coronal structures associated with spicular mass ejections and
heating of plasma to transition region and coronal
temperatures. Our observations suggest that a significant fraction of 
the highly dynamic loop fan environment associated with plage regions
 may be the result of the formation of such new coronal strands, a 
process that previously had been interpreted as the propagation of 
transient propagating coronal disturbances (PCD)s. Our observations
are supported by 2.5D radiative MHD simulations that show heating to
coronal temperatures in association with spicules. Our results suggest 
that heating and strong flows play an important role in maintaining
 the substructure of loop fans, in addition to the waves that permeate 
this low coronal environment.
\end{abstract}

\section{Introduction}
\label{sec:intro}

Chromospheric spicules are dynamic jet-like features that dominate the
solar limb and appear to penetrate the million-degree corona before
falling back to the surface. Their
nature has remained mysterious with many
explanations proposed for their origin \citep[for reviews, see][]{Sterling2000,Tsiropoula2012}. They
have long been considered as a plausible mechanism to provide plasma to
the corona \citep{Beckers1968,Pneuman1978,Athay1982}. 
The discussion about their role in the outer atmosphere was
recently revived, with the advent of Hinode \citep{Kosugi2007}, in particular
the Solar Optical Telescope \citep[SOT][]{Tsuneta2008} and the EUV Imaging
Spectrometer \citep[EIS][]{Culhane2007}, the Atmospheric Imaging
Assembly (AIA) onboard Solar Dynamics Observatory \citep{Lemen2012},
and the Interface Region Imaging Spectrograph
\citep[\iris][]{De-Pontieu2014}, as well as advanced 3D radiative MHD simulations
\citep[e.g.,][]{Martinez2011,Martinez2013}.

Observations with these spacecraft have provided a new view on
spicules, revealing the presence of: 1) relatively slow
($< 40$ km~s$^{-1}$), longer-lived (5-10 min) spicules that do not appear
to show signficant heating to transition region (TR) or coronal temperatures,
falling back towards the surface as chromospheric features
\cite[type I spicules, or dynamic fibrils when seen on the
disk, ][]{Hansteen2006}; 2) fast (40-100 km~s$^{-1}$) (type II) spicules that
are only briefly visible (1-2 min) in chromospheric observables
such as Ca II H 3968\AA\ \citep{De-Pontieu2007a,Pereira2012} and in which a fraction of the
plasma appears to be heated to at least TR temperatures
\citep{Pereira2014,Skogsrud2015, Rouppe2015} before returning to the
surface. We focus here on the impact of type
II spicules on the coronal mass and energy balance.

There have been several 
suggestions that these spicules play a significant role in
heating the corona. \citet{De-Pontieu2009a} suggested
that strong upflows of (multi)million-degree plasma 
seen at the footpoints of coronal loops as spectral line asymmetries \citep{Hara2008} 
were associated with upper chromospheric activity \citep[e.g.,][]{McIntosh2009}. 
\citet{De-Pontieu2011} suggested that the disk counterparts
of spicules \citep[rapid blueshifted events or
RBEs, see][]{Rouppe2009} are associated with brightenings in coronal
SDO/AIA lines in active regions \citep[for quiet Sun,
see][]{Henriques2016}. These studies have been the subject of
significant debate, both from an observational and a
theoretical point of view.  \citet{Madjarska2011} studied coronal hole
spicules at the limb and found no evidence for coronal counterparts
using lower resolution and lower signal-to-noise observations from SOHO/SUMER
\citep{Wilhelm1995}. \citet{Klimchuk2012} used simplified theoretical
considerations to reject a significant role of spicules in the coronal
heating issue, while \citet{Tripathi2013} and \citet{Patsourakos2014} 
studied spectral line asymmetries from EIS and argued that while spicules may play a role in
the coronal mass and energy balance, it is likely not a dominant
one. Lacking a theoretical model that captures the complexity of the
spicular environment, these studies are based on simplifying assumptions about
the physical scenario
focusing on single-field-line approaches that underestimate
the complexity of the spicular environment. The latter 
is observationally challenging to capture: they
rapidly fade in and out of various passbands, necessitating
a multi-instrument approach, and are so dynamic and finely structured
that many instruments do not resolve their
spatio-temporal evolution.

Here we attempt to address both issues. We exploit the \iris{} discovery of the TR counterparts
of spicules, in particular the fact that these are longer-lived than
the chromospheric spicules, allowing us to more easily track their
evolution. We use SDO/AIA to focus on the impact of spicules on coronal loops that
are connected to plage or enhanced network regions. These loops have long been known to be
permeated by propagating coronal disturbances (PCDs): rapid
($\sim 100$ km~s$^{-1}$) intensity disturbances whose exact nature
remains unknown. While it is clear that PCDs that originate from
sunspot umbrae are caused by sound waves, it is less clear whether
waves or flows cause these PCDs in plage regions, with both receiving observational and
theoretical support
\citep[e.g.][]{De-Moortel2002a,De-Moortel2002b,De-Pontieu2005,De-Moortel2009b,
  De-Pontieu2010,Verwichte2010,De-Moortel2012,Ofman2012,Tian2012,Wang2013,
 Petralia2014,De-Moortel2015,Samanta2015,Bryans2016}.
We also take advantage of recent developments in spicule modeling
\citep{Martinez2016,Martinez2017} that appear to show coronal heating associated
with spicules.


\section{Observations}
\label{sec:obs}

We use \iris{} slit-jaw observations of AR 12171 at
x$_{cen}$=464\arcsec, y$_{cen}$=-476\arcsec, 
taken on 26-Sep-2014 from 00:34-01:37 UTC using OBS-ID 3820107266. The \iris{} level 2 data was corrected
for dark current, flat-field, geometry and co-aligned as described in
\citet{De-Pontieu2014}. To boost signal-to-noise, the \iris{} data was
summed onboard 2x2, so that spatial pixels are 0.33\arcsec x
0.33\arcsec. Both 1330\AA\ and 1400\AA\ passbands were used which
are dominated by far-ultraviolet continuum (formed in the low
chromosphere) and, respectively, \ion{C}{2} 1335/1336\AA\ lines \citep[formed at
upper chromospheric and low TR temperatures, from 15,000-40,000 K][]{Rathore2015} and
\ion{Si}{4} 1394/1403\AA\ lines \citep[formed at TR temperatures from
20,000-300,000 K][]{Olluri2015}. 
Co-temporal \sdo/AIA observations in the 1600\AA, 171\AA\ and 193\AA\
passbands were prepped, coaligned, and normalized using the SolarSoft
{\tt aia\_prep} routine. The AIA data were interpolated in time and
space to match the \iris/SJI temporal cadence (10.4~s) and spatial
resolution (0.33\arcsec). 

We focus on a decayed plage region that is associated with ``plume''-like
coronal structures that emanate towards the south (Fig.~\ref{f1}).

\section{Results}
\label{sec:res}
\subsection{Observations}
\begin{figure*}[tbh]
\centering
\includegraphics[width=0.95\hsize]{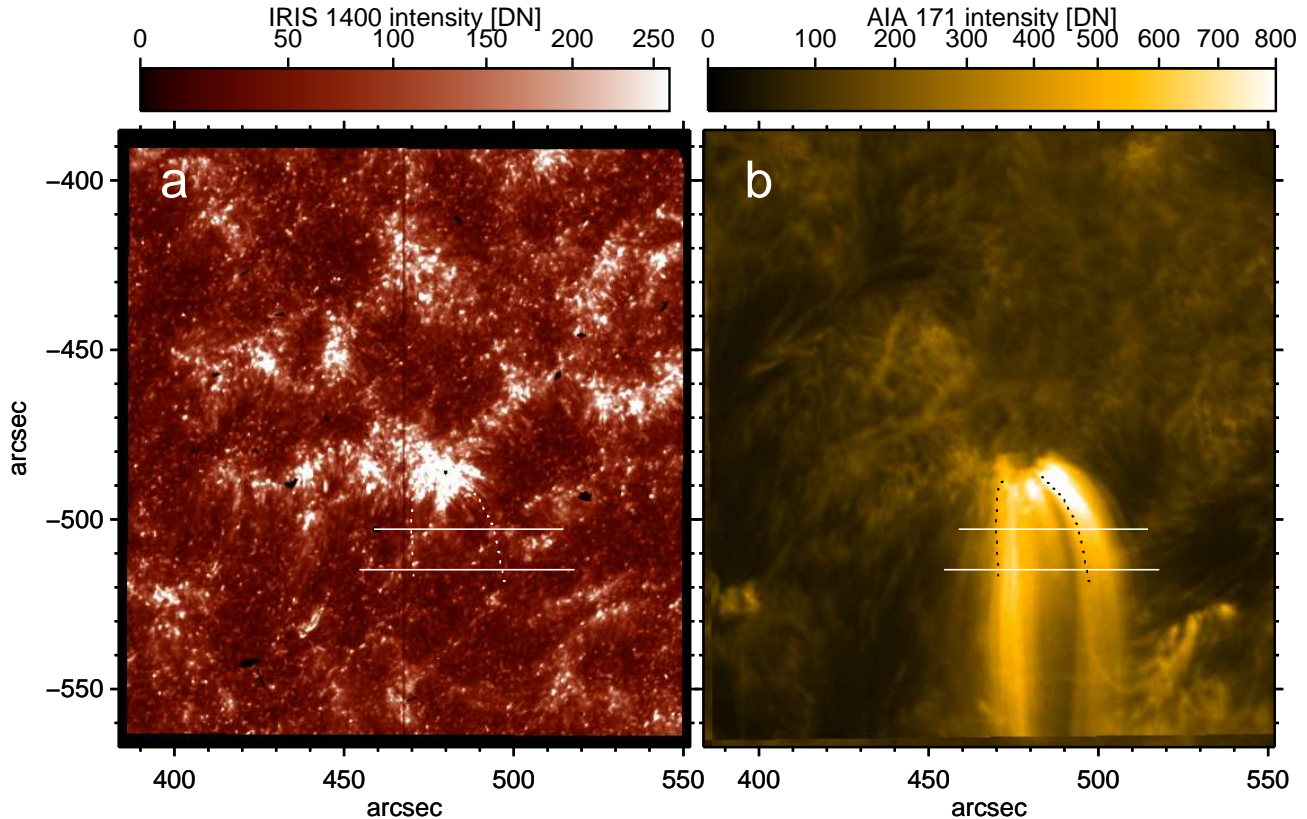}
\caption{\iris{} 1400\AA\ (a) and \sdo/AIA 171\AA{} (b) images taken
  on 26-Sep-2014 shown for context, with strands and cross-sections that are used for
  Figure~\ref{f4} overplotted. Accompanied by an online
animation that also shows \iris{} 1330\AA\ and the running difference
of \sdo/AIA 171\AA{}.}
\label{f1}
\end{figure*}

\begin{figure*}[tbh]
\begin{center}
\includegraphics[width=0.9\hsize]{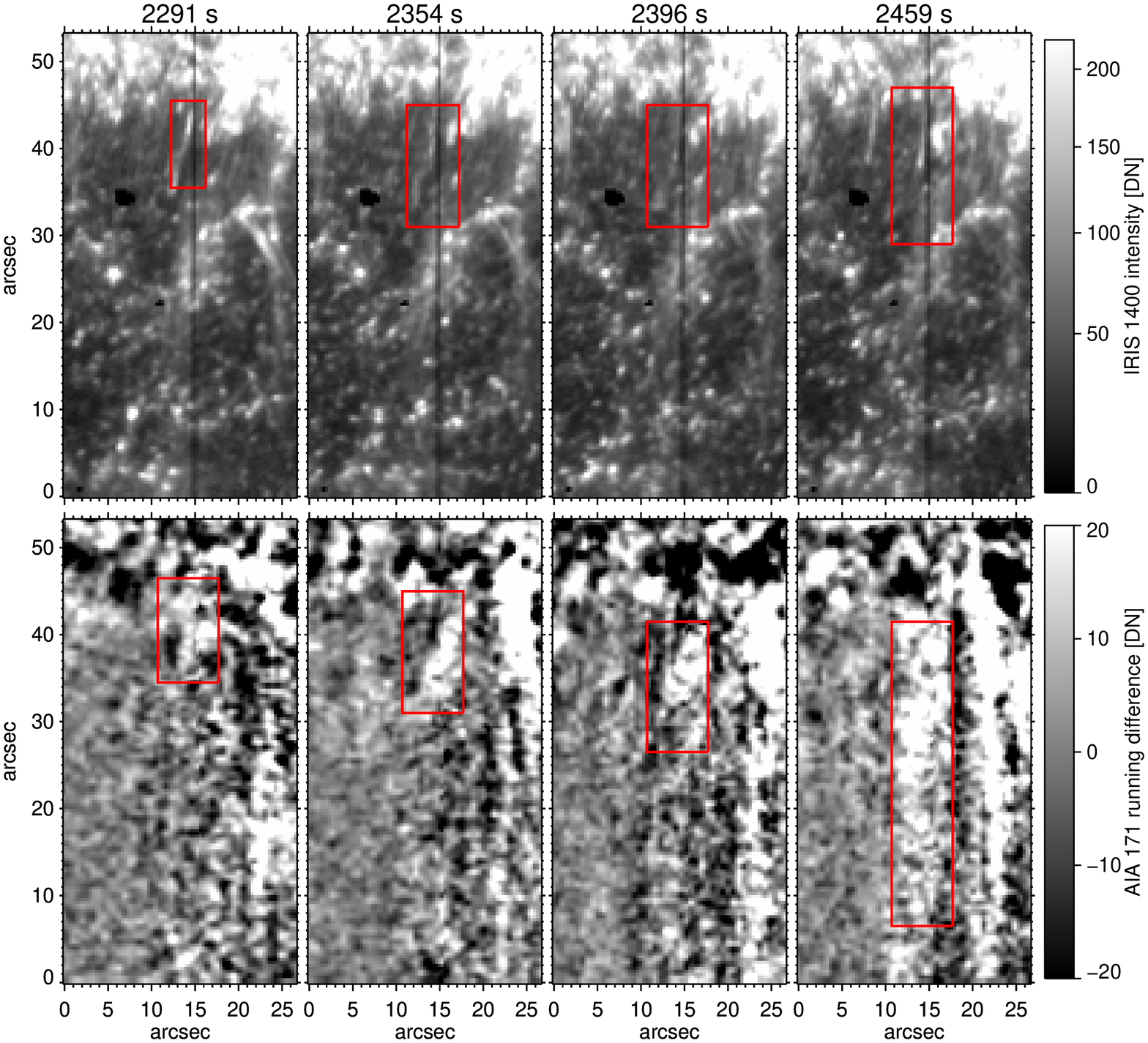}
\caption{Top: \iris{} 1400\AA\ images with the
  red boxes indicating the formation of the TR
  counterpart of a type II spicule. Bottom row: \sdo/AIA 171\AA{}
  running-difference images with red boxes indicating the
  formation and propagating of a PCD. This figure is accompanied by 
  two online animations that show the evolution of this event and a second, unrelated, event.} 
\label{f2}
\end{center}
\end{figure*}

\begin{figure*}[tbh]
\centering
\includegraphics[width=0.9\hsize]{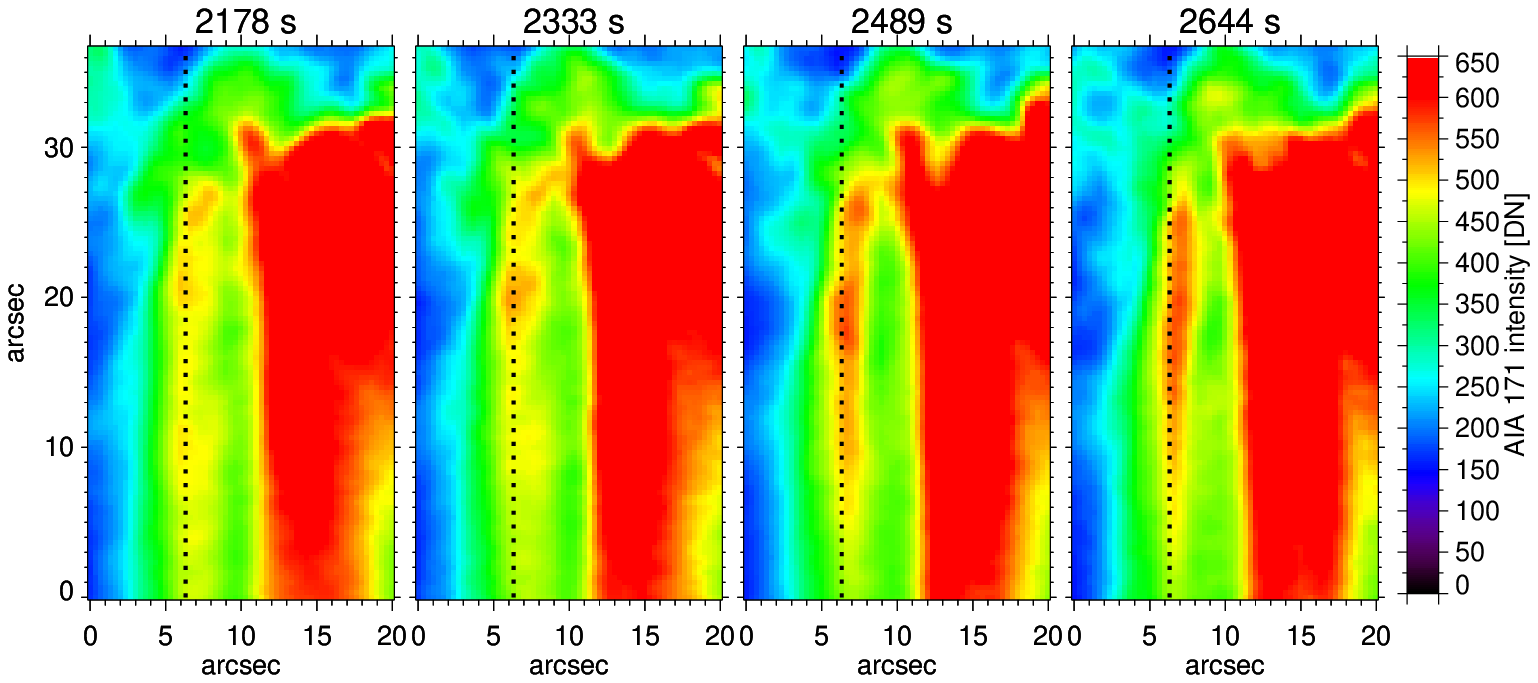}\\
\includegraphics[width=0.9\hsize]{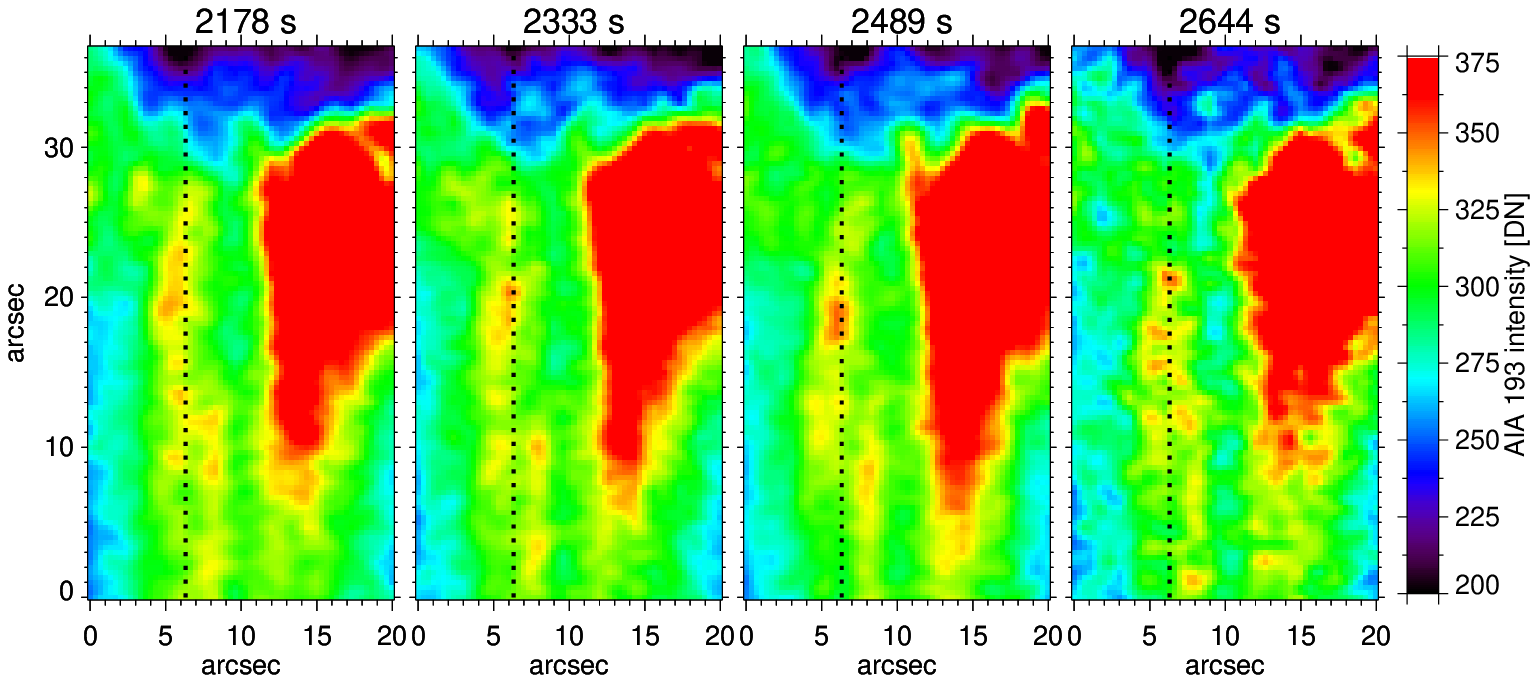}
\caption{Formation of a coronal loop strand in (top) \sdo/AIA
  171\AA{} and (bottom) 193\AA. The strand forms underneath and
  along the dashed vertical line, which is used for the space-time
  plot shown in the top of Fig.~\ref{f4}. Accompanied by a movie.}
\label{f3}
\end{figure*}

Movies of the \iris{} 1400\AA\ passband show that the footpoints of
these coronal loops are dominated by a multitude of spicule-like
features, which, given their appearance, are most likely caused by TR
emission (Si$^{3+}$ ions). These spicules originate
from magnetic flux concentrations, shooting away from the weak plage
at apparent speeds of 50-200 km~s$^{-1}$. One can often see the spicules
retract after reaching a maximum extent, although not always. They are
often not clearly visible along their whole length and fade as
they extend away from the plage, suggesting a complex thermal
environment and evolution, involving heating and/or cooling. 
The coronal loops rooted in the same region similarly 
show a lot of complexity, which traditionally has been associated with
``propagating coronal disturbances'' along pre-existing coronal
loops. However, closer inspection of
the timeseries associated with Fig.~\ref{f1} shows that much of this
activity is actually caused by a variety of coronal strands appearing and
disappearing, typically starting from the bottom of the loops,
followed by ``propagation'' away from the footpoint. It is also clear that the
PCDs that are observed here are not as cleanly periodic as
those associated with sunspot umbrae \citep[see also][]{De-Pontieu2010}). This is illustrated in
Figs.~\ref{f2}, \ref{f3} and accompanying movies.
 
The line-of-sight superposition and the multitude of events
overlapping in space and time means it is not straightforward to
disentangle individual events. However, analysis of the
evolution of several of the larger
spicules that stand out individually show an intriguing
connection between the \ion{C}{2} and \ion{Si}{4} spicules in
\iris{} slit-jaw movies and the formation of strands in the coronal
loop system (AIA). This is illustrated in Fig.~\ref{f2} and the accompanying
movies which show the temporal evolution of the event shown
in Fig.~\ref{f2} as well as a second event. In both cases we show the 
\ion{Si}{4} images (top row) and the running difference of the 171\AA\
AIA channel \--- calculated by differencing the current image
with that taken 62~s earlier, 
a commonly used method to
enhance the visibility of the intensity disturbances. For both
cases, we see the spicule form with apparent velocities in
the plane-of-the-sky of $\sim 50$ km~s$^{-1}$, accompanied by a brightening in 171\AA\ that
initially grows with the same apparent speed as the spicule. Towards the
time of maximum extension of the spicule ($t=2396$~s), the spicule seems
to stay roughly constant in length and we see (red rectangles in
Fig.~\ref{f2} and movies) that
the coronal counterpart grows rapidly to cover 30\arcsec\
in 50~s, suggesting an apparent speed of order 400 km~s$^{-1}$. A
similar evolution can be seen in the second movie accompanying Fig.~\ref{f2}:
that spicule grows with an apparent speed
of order 40 km~s$^{-1}$ until $t\sim1980~s$,
appears to stay at its maximum extent for a while, and then the
coronal counterpart grows another 30\arcsec\ in $\sim40$~s,
suggesting an apparent speed for this phase of $\sim 500$ km~s$^{-1}$. 

Figure~\ref{f3} shows how the coronal loop strand that
is associated with the spicule of Fig.~\ref{f2} becomes
visible in both AIA 171\AA\ and 193\AA\ passbands when using a
different color table that accentuates small intensity differences. The loop strand is more
clearly visible in 171\AA\ than in 193\AA\ suggesting that it reaches
temperatures closer to the formation temperature of \ion{Fe}{9} (log~$T = 5.9$) rather
than that of \ion{Fe}{12} (log~$T = 6.2$). This figure does not show the running
difference, but rather the original AIA intensity. It
illustrates how this particular feature is not a disturbance on top of
a pre-existing coronal loop structure, but the formation of
a completely new coronal strand. Detailed inspection of the AIA
timeseries 
shows that the formation of such strands
is not a rare event, but a common occurrence throughout this loop fan structure.

\begin{figure*}[tbh]
\centering
\includegraphics[width=0.95\hsize]{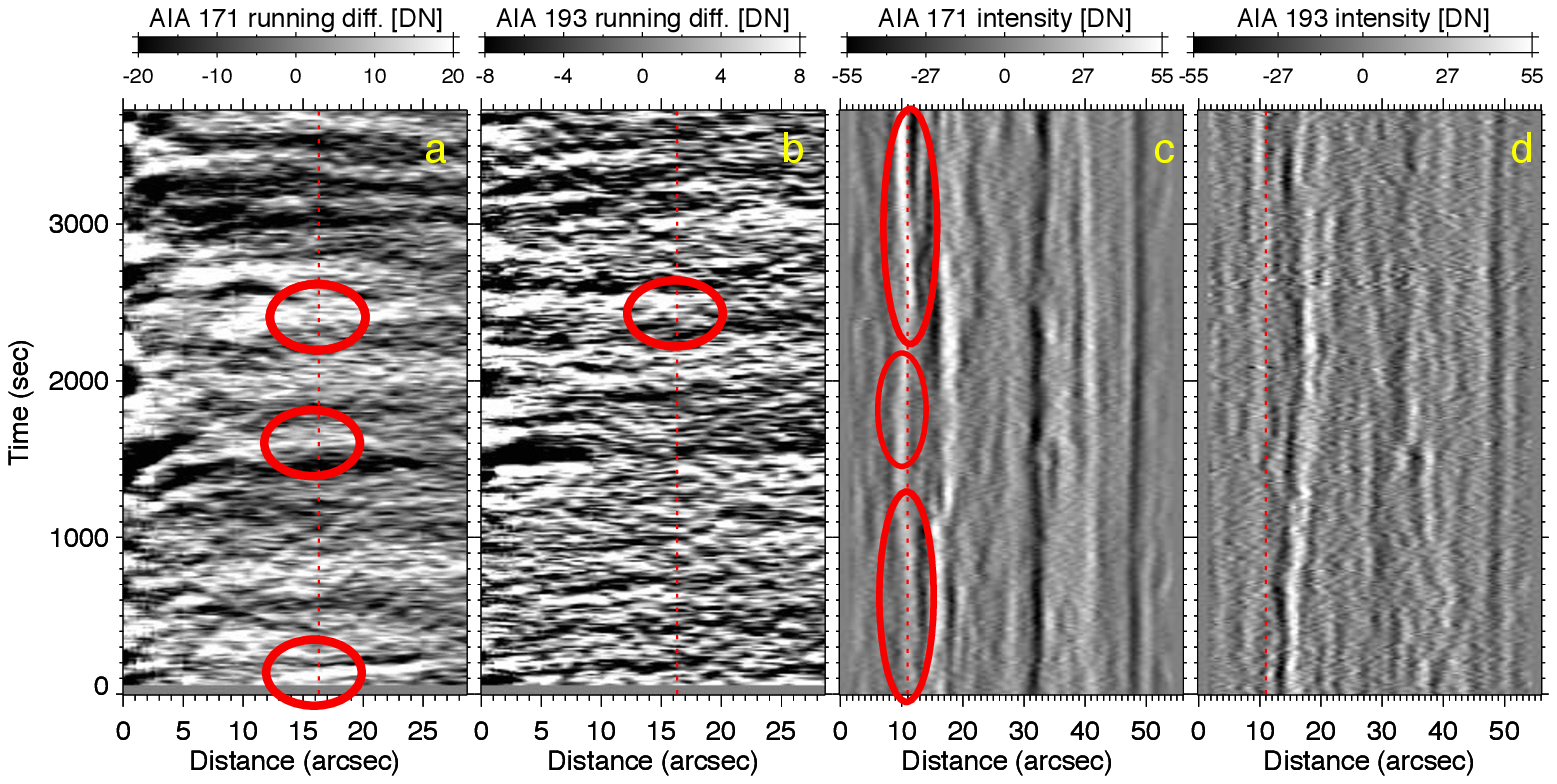}
\includegraphics[width=0.95\hsize]{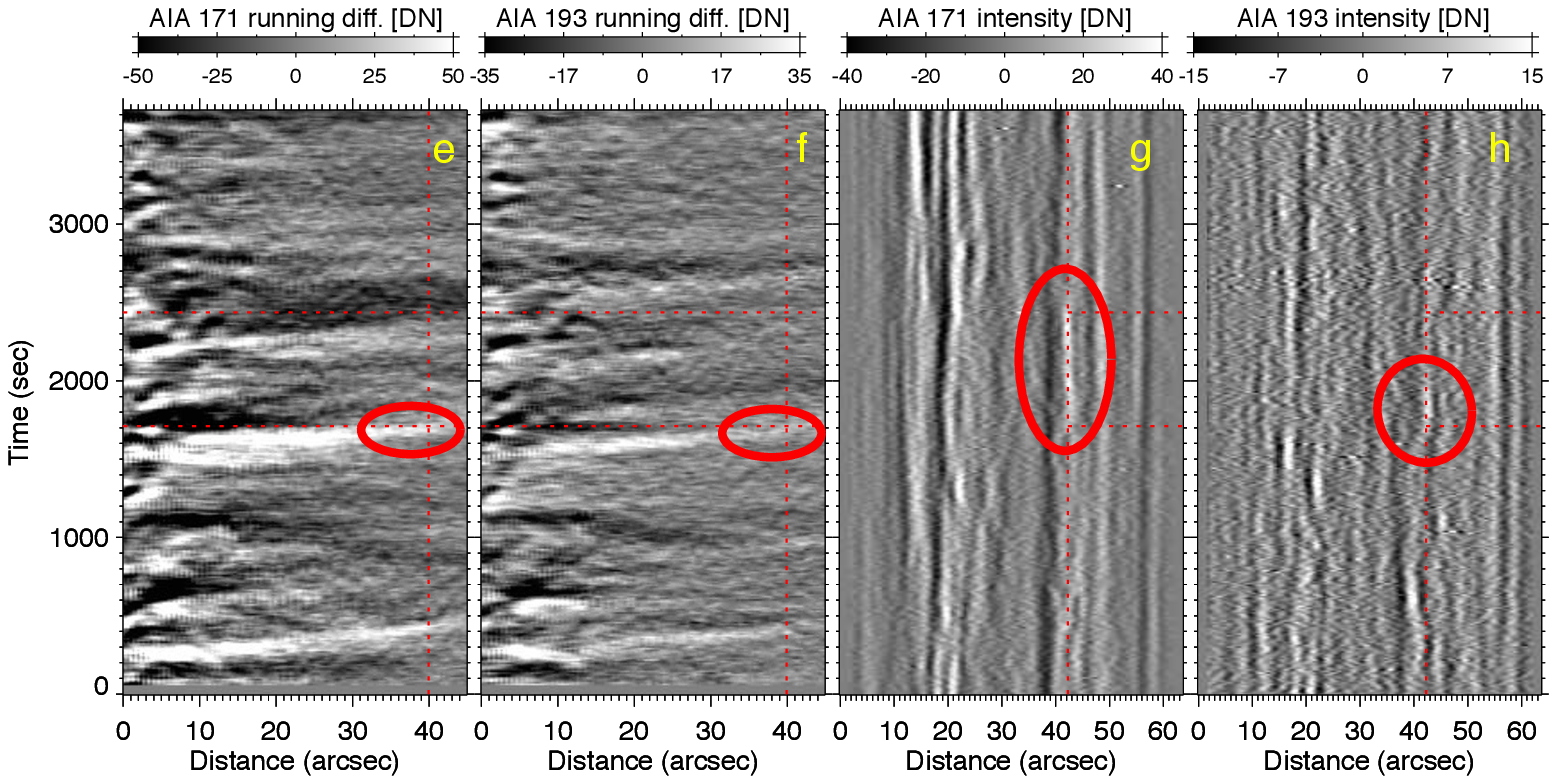}

\caption{Two examples (top and bottom) of strand formation
  associated with spicules. (A) Running difference of
  AIA 171\AA\ along the left vertical track in Fig.~\ref{f1} and the
  vertical line in Fig.~\ref{f3}, (B) same for
  193\AA, (C) Cross-section of unsharp masked AIA 171 \AA\ intensity along the upper horizontal
  line in Fig.~\ref{f1}. (D) Same for 193 \AA. Panels (E), (F), (G)
  and (H) show the same but for a different event, highlighted by the
  right vertical track and the bottom horizontal line
  in Fig.~\ref{f1}. Vertical lines in panels (A), (B), (E) and (F)
  indicate where the track crosses the horizontal cuts indicated in
  Fig~\ref{f1}. Horizontal lines and red circles identify specific events
  described in the text.}
\label{f4}
\end{figure*}

However, the formation of such a strand can also, deceptively, look
like a propagating coronal disturbance (PCD) as illustrated by
Fig.~\ref{f4}. The example in the top row shows the event that was
highlighted in Figs.~\ref{f2} and \ref{f3}. As shown in Fig.~\ref{f3}, this
event involved the formation of a new loop strand by $t=2200$~s, which
briefly appears as a propagating coronal disturbance in both 171 and
193\AA\ channels. Because of the nature of the running difference, the
longevity of the strand is not clear at all from panels (A) and
(B). This becomes much clearer in panels (C) and (D) which do not show
running difference and instead of the original 171 and 193\AA\ intensity, the intensity after
unsharp masking to enhance the small spatial scales of the coronal
loop strands. We show here
the spatio-temporal evolution for a cut (see upper horizontal line
in Fig.~\ref{f1}) across the loop strand that we
highlighted earlier. We see that this particular strand forms when the
PCD hits this location. While the PCD continues to propagate away from
this location, and the running difference plot suggests that a
``wave'' (perturbation) just passed through a background structure, panels (C) and
(D) indicate that the strand continues to exist long after the PCD has
left this region. This particular strand in fact undergoes repeated activity with many PCDs passing
through, and the strand being strengthened every time such a PCD passes by.

A more isolated case of strand formation associated with the passage of a PCD is
shown in the bottom of Fig.~\ref{f4} which reveals a 
short-lived passage of a PCD (at $t=1700$~s) that leads to the formation of a strand,
initially brightest in 211 \AA\ (not shown), followed by a brief event
in 193\AA\ (panel H) and then a prolonged presence in 171\AA. This
event occurs where the right hand track and bottom horizontal line
cross in Fig.~\ref{f1}. This
sequence of events strongly supports a scenario in which the PCD
appears to be associated not only with a
spicule, but also with heating of plasma to $\sim 1.5$ MK followed by apparently relatively rapid cooling and subsequent fading from the 171\AA\
passband after 10-15 minutes. Many more examples can be found in the
data that support this scenario of several phenomena associated with
spicules: triggering of PCDs, heating to coronal temperatures, and the
formation of loop strands.

\subsection{Simulations}
These results fit well with synthetic \ion{Fe}{9} 171\AA\
and \ion{Fe}{12} 193\AA\ observations from a 2.5D radiative MHD
simulation using Bifrost \citep{Gudiksen2011}, which captures
many physical processes important for the dynamics
and energetics of the solar atmosphere. This simulation covers a domain from the top of the convection
zone into the corona, including self-consistent chromospheric and coronal
heating \citep{Martinez2016,Martinez2017}. We 
included the effects of interactions between ions and neutrals, or ambipolar diffusion. Ambipolar diffusion plays a key role in the
formation of features that closely resemble type II
spicules, through a complex mechanism outlined
in \citet{Martinez2017}.  In summary, the interaction between weak,
granular-scale fields and strong flux concentrations leads to strong
magnetic tension, which can emerge into the chromosphere
through ambipolar diffusion and leads to a violent release of 
tension when the low plasma $\beta$ regime is reached in the middle to
upper chromosphere. This violent release leads to strong upward acceleration
of plasma, the formation of fast spicules and the generation of strong
transverse waves. In addition, currents, created through
several mechanisms including wave-mode coupling, gradients in
ambipolar diffusion and the interaction between emerging flux 
and pre-existing ambient field, are in part dissipated by ambipolar
diffusion in the spicule (leading to heating to TR
temperatures), and in part propagated into the corona at Alfv\'en speeds
where they lead to significant heating. 

\begin{figure*}[tbh]
\centering
\includegraphics[width=0.85\hsize]{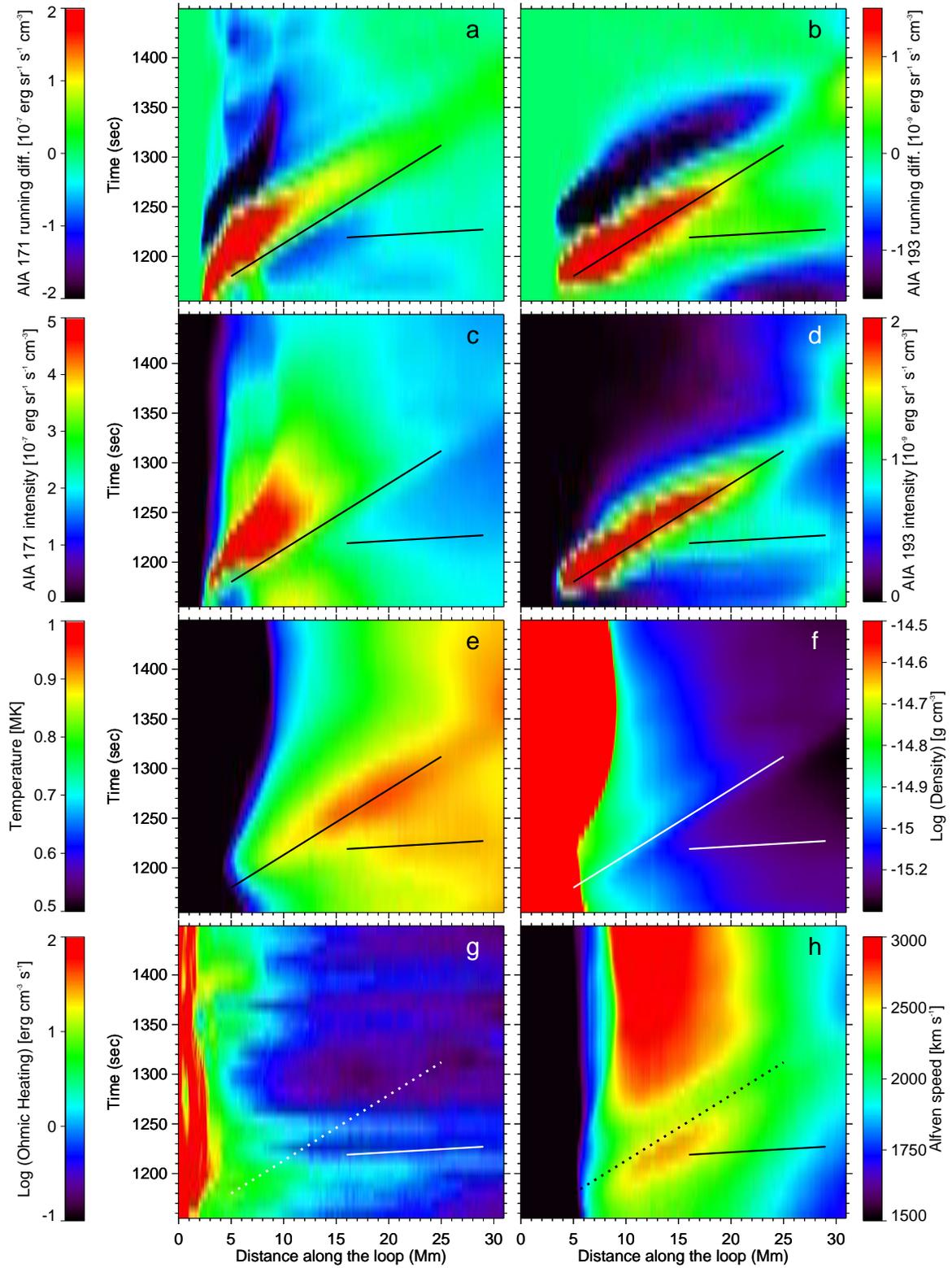}
\caption{Space-time plots from 2.5D MHD simulation of type II
  spicules showing (a) running difference intensity of \ion{Fe}{9}
  171\AA, (b) same for \ion{Fe}{12} 193\AA, (c) intensity of
  \ion{Fe}{9}, (d) same for \ion{Fe}{12}, (e) temperature, (f)
  density, (g) Ohmic heating, (h) Alfv\'en speed. Black lines show the
slopes of the apparent motion of the two different PCDs. This figure
is accompanied by an on-line animation.}
\label{f5}
\end{figure*}

The coronal impact of this scenario is illustrated
in Figure~\ref{f5}, which includes
running difference of 
\ion{Fe}{9} and \ion{Fe}{12} intensities along a track that covers a
spicule. While the \ion{Fe}{9} response is relatively simple
with the apparent propagation of a ``PCD'' at speeds of
$\sim150$~km~s$^{-1}$, the \ion{Fe}{12} running difference shows two
PCDs, one that is similar in slope as the \ion{Fe}{9}, and another
at much higher speeds ($\sim1600$~km~s$^{-1}$). Comparison with the
original intensity of both lines and various physical variables paints a complicated
picture and indicates that the interpretation of PCDs in terms
of physical mechanisms is not straightforward, as explained below. Both PCDs are causally linked to the launch of a fast
spicule. The chromospheric part of this spicule is visible as a
parabolic path in the temperature (e) which starts at the same time
and location as the \ion{Fe}{9} PCD. The cause of the \ion{Fe}{9} PCD
is, at low heights, a mix of flows associated with the TR and coronal counterparts of
the spicule (visible as a parabolic path that reaches distances of 10~Mm
in the original intensity of \ion{Fe}{9} (a) and in the density (f)), and the
coronal remnant of the shock wave that was involved in the spicule
acceleration. These flows are caused by the acceleration, compression
and heating of plasma associated with the spicule eruption. At greater heights ($>10$~Mm), the \ion{Fe}{9} PCD is mostly
determined by the shock-wave related disturbance \citep[similar
to][]{De-Pontieu2005,Petralia2014} since the coronal
counterpart of the spicule fades beyond distances of 12 Mm. The
\ion{Fe}{12} PCDs are even more complex: the slower \ion{Fe}{12} PCD
is a mix of spicular flows and shock-wave related
disturbances. However, the average slope that is drawn through the slow PCD in
this wavelength ignores the fact that the different physical
mechanisms lead to different slopes in the space-time plots, with the apparent speed increasing with
distance from the spicule footpoints. The \ion{Fe}{12} running
difference (b) also reveals a faster ``PCD'' that is caused by the rapid
formation of a loop strand. This loop strand is formed because of the
heating associated with the arrival and dissipation in the corona of the current that also heats
the spicular plasma, as well as thermal conduction that spreads the
released heat. This faster PCD at Alfv\'enic speeds could well be responsible for the rapid
propagation (400 km~s$^{-1}$) we see in our observations. The slope of the PCD with modest speed can also be 
affected by the heating from the currents (g).
The movie accompanying Fig.~\ref{f5} shows how the
currents that are created during the spicule formation
propagate rapidly into the corona (at Alfv\'enic speeds) and appear to
``meander'' through the coronal volume, similar to what is seen in
observations of coronal loop strands that often appear to ``move''
perpendicularly to their own axis. In our simulation, the
spicules are thus the harbingers of significant coronal heating both
through heating of spicular plasma and heating from current
dissipation associated with the spicule. The current dissipation in
the corona of our model occurs because of numerical resistivity
\citep{Gudiksen2011}.  In the solar atmosphere the
dissipation of this energy could be because of current dissipation on
small scales or dissipation of the Alfv\'en waves that are
triggered when the spicule is formed \citep{Martinez2017}.



\section{Discussion}
\label{sec:dis}

Our results support a scenario in which ``PCDs'' along
loops originating from plage or strong network regions are not
necessarily only a signature of magneto-acoustic waves, but often caused by a complex sequence
of events that involves generation of spicular flows and associated
shock waves that propagate into the corona, as well as plasma heating
through dissipation of electrical currents and magnetic waves. These currents are a key
component of the spicule formation which critically depends on ambipolar diffusion caused by the interaction
between ions and neutrals. Our observations provide a detailed view of how spicules, heated from
chromospheric to TR temperatures, set off PCDs, but also lead to the formation of new coronal
loop strands, thus locally contributing to the mass and energy balance
of the corona. Our results suggest that analysis
of PCDs through running differencing misses the fact that
plage-related loop strands are continuously formed and persist after
PCDs have ``passed''. Our results provide a natural explanation for
the often confusing reports of apparent speeds, which in our
simulations are caused by a mixture of real mass motions
of coronal plasma in response to spicular flows and heating, remnants
of shock waves generated during the spicule formation, heating through
spicule-associated currents, and subsequent thermal conduction. Our
simulations also show that idealized ``single-field line'' approaches
\citep[e.g.][]{Klimchuk2012} to spicule-associated coronal heating are
bound to fail: the spicular environment is highly complex, takes place
on many neighboring ``field lines'' some of which carry accelerated
plasma that is heated by ambipolar dissipation of electrical currents,
and others carry plasma that is heated by spicule-associated currents that rapidly
propagate into the corona. 

The complexity of physical
mechanisms in these simulations highlights why it
is so difficult to determine the ``spicule contribution to coronal
heating''. Such an endeavor is driven by an approach that
is based on observational phenomena (``chromospheric'' spicules)
which our simulated (and likely solar) reality defies: spicules are neither
chromospheric, TR or coronal phenomena; they are all of
the above, and their dynamics and energetics are intimately tied to
that of the corona. Our results indicate that the
currents and waves associated with
spicule formation should not be ignored in future studies of coronal heating.





\longacknowledgment{}
%


\end{document}